\documentclass[pdflatex,sn-mathphys-num]{sn-jnl}% Math and Physical Sciences Numbered Reference Style
\usepackage{graphicx}% Include figure files
\graphicspath{
    {.../Fig/300ppi/}
}%
\usepackage{multirow}%
\usepackage{amsmath,amssymb,amsfonts}%
\usepackage{amsthm}%
\usepackage{mathrsfs}%
\usepackage[title]{appendix}%
\usepackage{xcolor}%
\usepackage{textcomp}%
\usepackage{manyfoot}%
\usepackage{booktabs}%
\usepackage{algorithm}%
\usepackage{algorithmicx}%
\usepackage{algpseudocode}%
\usepackage{listings}%
\usepackage{kotex}
\usepackage{dcolumn}% Align table columns on decimal point
\usepackage{bm}% bold math
\usepackage{float}
\renewcommand{\vec}[1]{{\bm{\mathrm{#1}}}}

\begin{document}

\title[Article Title]{Gauge-Field-Mediated Symmetry Breaking of Matters Under Electromagnetic Fields and Its Impact on Spin Dynamics}


\author[1]{\fnm{Uiseok} \sur{Jeong}}\email{uiseok95@unist.ac.kr}
\author[1]{\fnm{Esmaeil} \sur{Taghizadeh Sisakht}} \email{sisakht@unist.ac.kr}
\author[2]{\fnm{Angel} \sur{Rubio}} \email{angel.rubio@mpsd.mpg.de}
\author[3]{\fnm{Carsten} \sur{A. Ullrich}} \email{ullrichc@missouri.edu}
\author*[4]{\fnm{Kyoung-Whan} \sur{Kim}}\email{kwkim@yonsei.ac.kr}
\author*[1,2]{\fnm{Noejung} \sur{Park}}\email{noejung@unist.ac.kr}

\affil[1]{\orgdiv{Department of Physics}, \orgname{Ulsan National Institute of Science and Technology}, \orgaddress{\street{50, UNIST-gil}, \city{Ulju-gun}, \postcode{44919}, \state{Ulsan}, \country{Republic of Korea}}}

\affil[2]{\orgdiv{Department of Physics}, \orgname{Max Planck Institute for the Structure and Dynamics of Matter}, \orgaddress{\street{Luruper Ch 149}, \city{Hamburg}, \postcode{22607}, \state{Hamburg}, \country{Germany}}}

\affil[3]{\orgdiv{Department of Physics and Astronomy}, \orgname{University of Missouri}, \orgaddress{\city{Columbia}, \postcode{65211}, \state{Missouri}, \country{USA}}}

\affil[4]{\orgdiv{Department of Physics}, \orgname{Yeonsei University}, \orgaddress{\street{50, Yonsei-ro}, \city{Seodaemun-gu}, \postcode{03722}, \state{Seoul}, \country{Republic of Korea}}}


\abstract{When a condensed-matter system is subjected to external electromagnetic fields, the gauge-invariant formulation of physical operators must explicitly incorporate the gauge-field contribution. However, in the context of spin–orbit coupling (SOC), this gauge-field term is often regarded as negligible or merely additive compared to the canonical SOC, which is typically localized near atomic cores. Here, we demonstrate that the symmetry breaking and consequent spin dynamics are governed by the gauge-field term, without which the spins remain symmetry-constrained. We perform real-time time-dependent density functional theory calculations to investigate spin–orbit dynamics, focusing on representative cases with mirror, glide, and screw-rotational symmetry. We demonstrate that when the gauge-field term in the time-dependent Hamiltonian perturbs the symmetry of the canonical term, a dynamical spin state gradually develops during the time evolution, beyond the symmetry-frozen states. We suggest that, for nonequilibrium spin–orbit dynamics, the gauge-invariant formulation of SOC is not only formally required but also quantitatively essential, even for a weak external field.}


\maketitle

\section{Introduction}\label{sec1}
Investigating spin–orbit–coupled electron dynamics in condensed-matter systems driven by external electromagnetic fields has become a timely and central problem, underpinning ultrafast and terahertz spintronics as well as light-field control of spin and orbital angular momentum\cite{RN1, RN2, RN3, RN4, RevModPhys.82.2731, RevModPhys.91.035004, LIU2023100705}. Spin–orbit dynamics in solids is primarily governed by the spin–orbit coupling (SOC) Hamiltonian, which arises as a leading relativistic correction that links the electron’s spin and orbital degrees of freedom\cite{PhysRev.78.29, PhysRevB.79.094422}. When electrons in condensed matter are subjected to external electromagnetic fields, the electronic Hamiltonian necessarily involves gauge-dependent electromagnetic potentials\cite{Tokatly_2010, PhysRevB.81.125123, TATARA2019208}.The corresponding SOC Hamiltonian and the gauge-covariant equations of motion can be derived naturally from the Lorentz- and gauge-invariant theory for the coupling between the Maxwell and Dirac fields. A perturbative expansion of this theory in the powers  $(v/c)^2$ can be summarized as follows\cite{Peskin:1995ev, shankar2004principles, PLB2010, PhysRevA.101.022501, PhysRevA.102.022804, PhysRevA.106.022816, PRXQuantum.5.020322, RN147, doi:10.1142/S021773231550128X}:
\begin{equation}
\begin{gathered}
  {\mathcal{L}_{{\text{eff}}}}[{\psi ^\dagger }(x),\psi (x)] \\ 
   = {\psi ^\dagger }\left( {i{D^0} + \frac{{{{\vec{D}}^2}}}{{2m}} + {c_F}\frac{e}{{2m}}{\vec{\sigma }} \cdot {\vec{B}}} \right)\psi  \\ 
   + {\psi ^\dagger }\left( {\frac{{{{\vec{D}}^4}}}{{8{m^3}{c^2}}} - {c_D}\frac{e}{{8{m^2}{c^2}}}\left( {{\vec{D}} \cdot {\vec{E}} - {\vec{E}} \cdot {\vec{D}}} \right) - {c_S}\frac{{ie}}{{8{m^2}{c^2}}}{\vec{\sigma }} \cdot \left( {{\vec{D}} \times {\vec{E}} - {\vec{E}} \times {\vec{D}}} \right) +  \cdots } \right)\psi  \\ 
   - \frac{{{d_1}}}{4}{F_{\mu \nu }}{F^{\mu \nu }} + \frac{{{d_2}}}{{{m^2}}}{F_{\mu \nu }}{F^{\mu \nu }} +  \cdots . \\ 
\end{gathered}
\label{Eq1}
\end{equation}
In Eq. (\ref{Eq1}), the covariant derivative is defined as ${D^\mu } = {\partial ^\mu } + i(e/c){A^\mu }$, and in what follows, we denote the electron charge as $-e$. Up to the order of $(v/c)^2$, three corrections arise beyond the non-relativistic Schrödinger theory\cite{shankar2004principles}: the relativistic kinetic energy correction(${\frac{{{{\vec{D}}^4}}}{{8{m^3}{c^2}}}}$), the SOC term($\frac{{ie}}{{8{m^2}{c^2}}}{\vec{\sigma }} \cdot \left( {{\vec{D}} \times {\vec{E}} - {\vec{E}} \times {\vec{D}}} \right)$), and the Darwin term(${c_D}\frac{e}{{8{m^2}{c^2}}}\left( {{\vec{D}} \cdot {\vec{E}} - {\vec{E}} \cdot {\vec{D}}} \right)$). In many areas of condensed matter studies, particularly for magnetism, spin textures, and topological states, the SOC term has been a central ingredient. In this spirit, the SOC contribution to the time-dependent Schrödinger-Pauli equation must be written as\cite{RevModPhys.65.733}
\begin{equation}
\begin{gathered}
{\hat H_{\text{SOC}}} =  - \frac{{ie\hbar }}{{8{m^2}{c^2}}}{\vec{\sigma }} \cdot \left( {{\vec{\hat \pi }} \times {\vec{E}} - {\vec{E}} \times {\vec{\hat \pi }}}\right) \\
= i\frac{{e\hbar }}{{8{m^2}{c^2}}}\frac{\partial }{{\partial t}}{\vec{\sigma }} \cdot \left( {\vec \nabla  \times {\vec{A}}} \right) + \frac{{e\hbar }}{{4{m^2}{c^2}}}{\vec{\sigma }} \cdot (\vec{E} \times \vec{\hat{\pi}})    
\end{gathered}
\label{Eq2}
\end{equation}
In Eq. (\ref{Eq2}), the gauge-invariant mechanical momentum is defined as ${\vec{\hat \pi }} = {\vec{\hat p}} + (e/c){\vec{A}}$. Hereafter, we explicitly denote the external source of electromagnetic with the subscript as $\vec A_{\text{ext}}(t)$. When the magnetic field is absent or is merely static and constant, $(\partial /\partial t)\nabla  \times {\vec A_{\text{ext}}} = 0$, the SOC part of the Hamiltonian, as written in Eq. (\ref{Eq2}), can be summarized as the following form:
\begin{equation}
\begin{gathered}
  {{\hat H}_{\text{SOC}}} = \frac{{\hbar e}}{{4{m^2}{c^2}}}{\vec{\sigma }} \cdot \left( {{\vec{E}} \times {\vec{\hat \pi }}}\right) = {{\hat H}_{\text{SOC}}^{\text{CN}}} + {{\hat H}_{\text{SOC}}^{{\text{GF}}}}, \\ 
  {{\hat H}_{\text{SOC}}^{\text{CN}}} = \frac{\hbar }{{4{m^2}{c^2}}}{\vec{\sigma }} \cdot( \vec E(\vec{r},t) \times ( - i\hbar \vec \nabla )), \\ 
  {{\hat H}_{\text{SOC}}^{{\text{GF}}}} = \frac{{\hbar e}}{{4{m^2}{c^3}}}{\vec{\sigma }} \cdot( \vec E(\vec{r},t) \times {{\vec{A}}_{\text{ext}}}(t)). \\ 
\end{gathered}
\label{Eq3}
\end{equation}
In Eq. (\ref{Eq3}), and throughout the present work, we label the SOC components according to the momentum contributions: the canonical part and the gauge-field part, denoted as ${\hat H_{\text{SOC}}^{\text{CN}}}$ and ${\hat H_{\text{SOC}}^{\text{GF}}}$, respectively. In early studies, the effect of external electromagnetic fields was often assumed to be negligible compared to the canonical spin–orbit coupling term (${\hat H_{\text{SOC}}^{\text{CN}}}$), in which the gradient of a scalar potential--strongly localized near atomic cores--dominates over externally applied fields\cite{krieger2014laser, DEWHURST201692, q46t-hck1}. More recently, the gauge-field part has been reexamined in the context of $U(1) \times SU(2)$ gauge invariance in the formulation of relativistic density functional potentials and the SOC Hamiltonian\cite{RN258, PhysRevB.102.081121, PhysRevB.96.035141}. In this context, particular attention has been paid to the definition of the charge current operator and its impact on charge dynamics and the consequent high-harmonic generation. 

In the present work, we focus on symmetry breaking driven by an external field. We show that, for a solid-state system in the velocity gauge, the symmetry breaking of the SOC Hamiltonian arises exclusively from the gauge-field term, underscoring its far-dominant role compared to previously known effects. Through a few real-materials real-time Time-Dependent Density Functional Theory(rt-TDDFT) calculation, we demonstrate that, no matter how weak the external field may be ($ - c{(\partial{\vec{A}}_{\text{ext}}(t) /\partial t)} = {{\vec{E}}_{\text{ext}}}(t)$), the effect of ${\hat H_{\text{SOC}}^{\text{GF}}}$ cannot be ignored. In particular, when it possesses a symmetry distinct from the canonical term, the dynamical spin state can be significantly affected, releasing the symmetry constraint. Even if the strength of the external field is weak, in time-dependent equations, the effect accumulates over long time scales and thus becomes non-negligible.

To establish a clear connection between symmetry, gauge invariance, and real-time spin dynamics, we adopt a two-pronged strategy. We first examine the Edelstein-type spin dynamics of a two-dimensional band system driven by an external field. We analyze a minimal two-band model of the Rashba-type SOC in a controlled real-time framework. And then, we perform the rt-TDDFT calculations for a SOC system with mirror or glide symmetries, together with one-dimensional screw-rotational symmetry. We quantitatively studied how the symmetry-breaking effect of the gauge field term characterized the spin dynamics. In the last part, we quantitatively analyze the features of symmetry breaking by the gauge field term.

\section{Results}\label{sec2}
\begin{figure}[h]
    \includegraphics{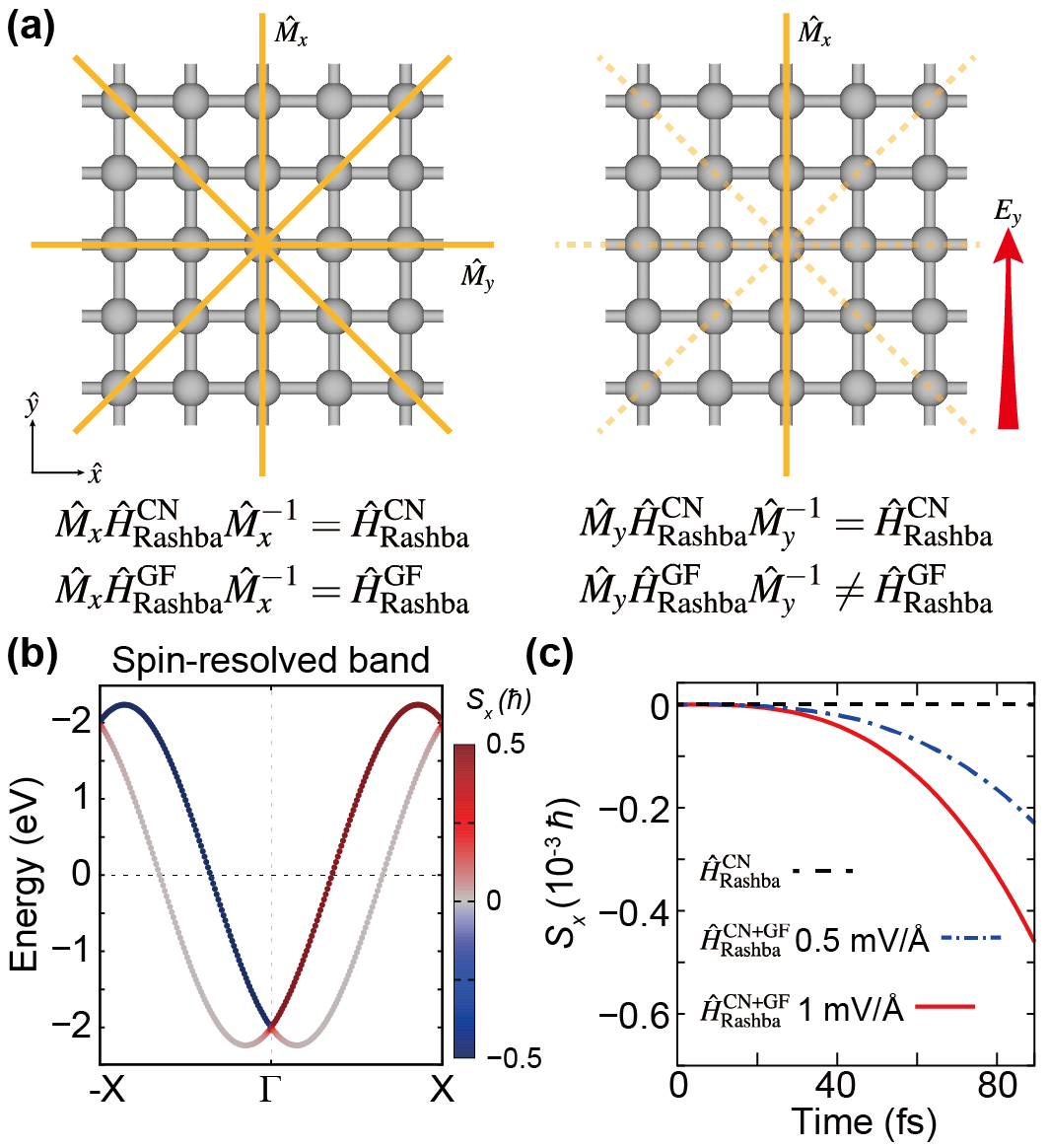}
    \centering
    \caption{Mirror-reflection properties of the canonical and gauge-field parts of the SOC Hamiltonian for a two-dimensional square lattice. (a) The yellow lines denote the mirror planes, specifically $\hat{M}_x$ and $\hat{M}_y$, which are perpendicular to the $y$ and $x$ axes, respectively. When an external field is applied along the $y$ direction, $\vec{A}_{\text{ext}}(t)=(0, A_{\text{ext}}(t), 0)$, $\hat M_y$ mirror is broken, and the gauge-field part of the SOC Hamiltonian becomes asymmetric with respect to $M_x$. (b) Static spin-projected band structure in which the $S_x$ value measured for each band eigenstate is denoted by color. (c) Time-dependent spin value $S_x(t)$ obtained from the real-time time-dependent Schrodinger equation, either with only canonical SOC or including the gauge-field part. Results are shown for two field strengths (0.5 mV/\AA ~and 1.0 mV/\AA). }
    \label{Fig1}
\end{figure}
\textit{Two-band Rashba model of 2D square lattice}---As an example of a minimal two-band Rashba model, here we examine a 2D square lattice, as summarized in Fig. \ref{Fig1}(a). To have a quantitative comparison between the canonical SOC term and the gauge-field induced corrections, the components of k-resolved Hamiltonian are denoted as follows:
\begin{equation}
\begin{gathered}
  \hat H({\vec{k}}) = {{\hat H}_0}({\vec{k}}) + {{\hat H}^{\text{CN}}_{{\text{Rashba}}}}({\vec{k}}), \\ 
  {{\hat H}_0}({\vec{k}}) =  - 2t\left( {\cos ({k_x}) + \cos ({k_y})} \right) \hat I, \\ 
  \hat H_{{\text{Rashba}}}^{{\text{CN}}}({\vec{k}}) = 2\alpha \left[ {\sin ({k_y}){{\hat \sigma }_x} - \sin ({k_x}){{\hat \sigma }_y}} \right].\\
\end{gathered}
\label{Eq4}
\end{equation}
Note that in this Rashba SOC term, as written in the third line of Eq. (\ref{Eq4}), is derived by substituting a constant perpendicular electric field($E_{\text{ext}}\hat z$) for $\vec{E}(\vec{r},t)$ in the canonical part of SOC in Eq. (\ref{Eq3}). The static energy band structures are summarized in Fig. \ref{Fig1}(b), which shows the clear spin texture and typical Rashba-type band splitting. The spin texture of these bands in the 2-dimensional Brillouin zone is summarized in Supplementary Fig. 1 in the Supplementary Information. We now consider the effect of mirror reflection symmetry. For k-resolved 2$\times$2 representation of the Hamiltonian, the mirror reflection operation flips the perpendicular component of the momentum and the in-plane components of the spin as a defining nature of the polar vector and the axial vector, respectively, as explicitly summarized in Eq. (\ref{Eq5}).
\begin{equation}
\begin{gathered}
{\hat{M}}_y:(k_x,k_y)\to(k_x,-k_y), \\
{\hat{M}}_y:(\sigma_x,\sigma_y,\sigma_z)\to(-\sigma_x,\sigma_y,-\sigma_z),\\
{\hat{M}}_y:S_x \to -S_x.
\end{gathered}
\label{Eq5}
\end{equation}
Note that the canonical part of the Rashba-SOC Hamiltonian is invariant under the mirror reflection ($\hat M_y$ as well as $\hat M_x$). Now, we consider the time-dependent situation. We examine how the mirror reflection symmetry is altered by the external electric bias depicted in the right panel of Fig. \ref{Fig1}(a). Accommodating the electric bias $\vec{E}_{\text{ext}}(t)=E_{\text{ext}}(t)\hat{y}$, the time-dependent Hamiltonian can be constructed by Peierls substitution:
\begin{equation}
    \begin{gathered}
  \hat H(\vec{k}(t)) = {\hat H_0}(\vec{k}(t)) + \hat{H}_{{\text{Rashba}}}^{{\text{CN+GF}}}(\vec{k}(t)),\\
  \hat{H}_{{\text{Rashba}}}^{{\text{CN+GF}}}(\vec{k}(t))=2\alpha \left[ {\sin ({k_y}(t)){{\hat \sigma }_x} - \sin ({k_x}(t)){{\hat \sigma }_y}} \right]\\
  \vec{k}(t)=\vec{k}_0+e\vec{A}_{\text{ext}}(t)/\hbar c,\\
  {{\vec{E}}_{\text{ext}}}(t) = - \frac{1}{c}\frac{\partial }{{\partial t}}{{\vec{A}}_{\text{ext}}}(t).
  \end{gathered}
  \label{Eq6}
\end{equation}
Here, we emphasize that, in the presence of $\vec{A}_{\text{ext}}(t)$, the SOC Hamiltonian($\hat{H}_{{\text{Rashba}}}^{{\text{CN+GF}}}$) is no longer symmetric under the mirror reflection($\hat M _y$), as considered in Eq. (\ref{Eq5}). We explicitly calculated the time-dependent Schrodinger equation and obtained a real-time profile of spin expectation values.

\begin{equation}
\begin{gathered}
    i\hbar \frac{\partial }{{\partial t}}\left|{\psi _{n,\vec{k}}}(t)\right\rangle = \hat H(\vec{k}(t))\left|{\psi _{n,\vec{k}}}(t)\right\rangle, \\ 
    {\vec{S}}(t) = \sum\limits_{n,{\vec{k}}} {{f_{n,{\vec{k}}}}\frac{\hbar}{2}\left\langle {{\psi _{n,{\vec{k}}}}(t)\left| {{\vec{\hat \sigma}}} \right|{\psi _{n,{\vec{k}}}}(t)} \right\rangle },
\end{gathered}
\label{Eq7}
\end{equation}
where the $f_{n,\vec{k}}$ in Eq. (\ref{Eq7}) denotes the occupation of the band states. The time-evolution of Rashba spin profile($S_x(t)$) is plotted in Fig. \ref{Fig1}(c), which obviously proves that the effect of the gauge-field part of the SOC $\hat H^{\text{CN+GF}}_{\text {Rashba}}$. Note that, when we exclude the gauge-field contribution, considering only the canonical part of the SOC, the mirror $\hat M_y$ is preserved, and the  $S_x(t)$ is still frozen under the symmetry constraint.

\begin{figure}[h]
    \includegraphics{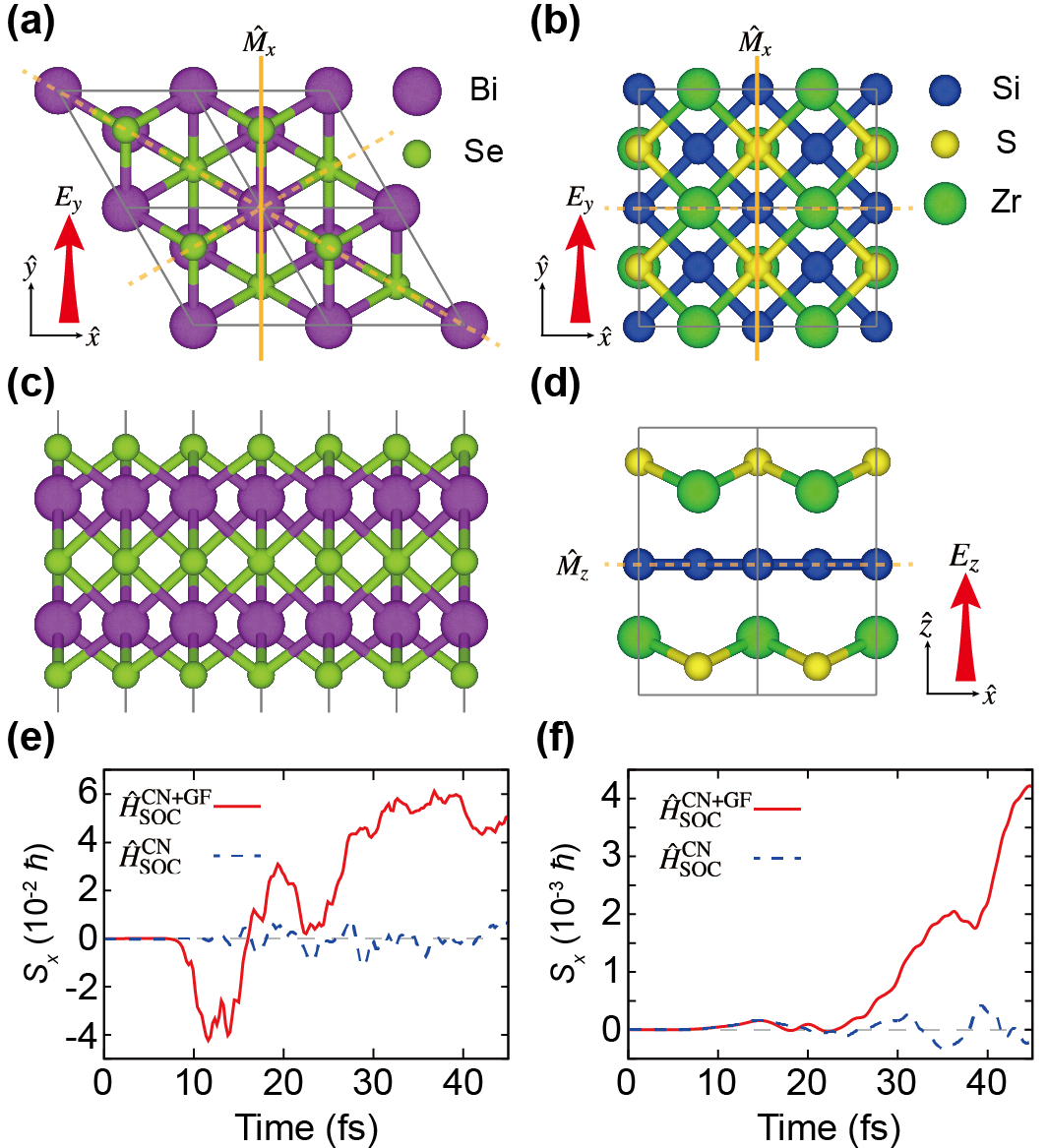}
    \centering
    \caption{Symmetry breaking induced by external fields and the consequent spin dynamics in Bi$_2$Se$_3$ and ZrSiS. (a, b) Top views and (c, d) side views of Bi$_2$Se$_3$ and ZrSiS, respectively. The mirror planes of each structure are indicated as yellow solid and dashed lines. (c, d) Time profile of $S_x (t)$, obtained from rt-TDDFT for (c) Bi$_2$Se$_3$ and (d) ZrSiS.
    }
    \label{Fig2}
\end{figure}

We now perform \textit{ab initio} rt-TDDFT calculations to investigate the spin dynamics in real materials either with $\hat H_{{\text{SOC}}}^{{\text{CN}}}$ or $\hat H_{{\text{SOC}}}^{{\text{CN+GF}}}$(read Eq. (\ref{Eq13}) of Method section). As in the model Hamiltonian analysis above, we focus here on how the symmetry of the material is modified by the external field. As summarized in Supplementary Note 1, the canonical component of the SOC is implemented via the standard pseudopotential generation scheme: the dominant part of $\vec E(\vec r,t)$ in Eq. (\ref{Eq3}) is sharply localized at the atomic core as the gradient of a spherical symmetric potential, and the corresponding SOC is provided through the form of non-local projectors, as implemented in standard pseudopotential packages. We first consider Bi$_2$Se$_3$, which is widely known for its substantial Rashba-Edelstein effect\cite{RN254, RN256, PhysRevB.103.035412}. The two-dimensional plane of the monolayer Bi$_2$Se$_3$ possesses $C_{3v}$ symmetry, and the corresponding 3 mirror planes are denoted by yellow dashed and solid lines in Fig. \ref{Fig2}(a). Owing to the axial-vector nature of spin, its expectation value is oriented perpendicular to each mirror plane. Consequently, the presence of multiple mirror symmetries enforces the spin expectation value to vanish.
It is noteworthy that the spin computed with $\hat H_{{\text{SOC}}}^{{\text{CN}}}$ remains frozen under the symmetry constraint, as presented by dashed line in Fig. \ref{Fig2}(e), in stark contrast to the substantial emergence of $S_x (t)$ obtained with $\hat H_{{\text{SOC}}}^{{\text{CN+GF}}}$. For both methods--whether the $\hat H_{{\text{SOC}}}^{{\text{CN}}}$ or $\hat H_{{\text{SOC}}}^{{\text{CN+GF}}}$--$S_y(t)$ and $S_z(t)$ remain negligible, at the level of numerical noise (see Supplementary Fig. 4 in Supplementary Information), consistent with the aforementioned axial-vector nature of the spins.

As an additional example, we investigate ZrSiS, which is widely known for Dirac semimetal electronic structure\cite{snby-9xsr, Lodge_2024, PhysRevB.101.064430, article}. In equilibrium, in addition to the two mirror symmetries $\hat M _x$ and $\hat M _y$(see Fig. \ref{Fig2}(b)), this material possesses a nonsymmorphic glide symmetry. Specifically, a mirror operation with respect to the $z$-direction is combined with a fractional translation along the $x$- and $y$-axes (see Fig. \ref{Fig2}(d)), leading to the gliding operation.
\begin{equation}
    \hat G = {\hat M_z}\hat T(\frac{a}{4}\hat x + \frac{a}{4}\hat y) = \hat T(\frac{a}{4}\hat x + \frac{a}{4}\hat y){\hat M_z}
    \label{Eq9}
\end{equation}
Because the spin operator is unaffected by the fractional translation, the spin transformation under the glide operation will be given by
\begin{equation}
    {\hat G^{-1} }{\mathbf{\hat S}}\hat G = {\hat M_z^{-1}}{\mathbf{\hat S}}{\hat M_z}.
\end{equation}
In other words, due to the axial-vector nature of spin, the glide symmetry enforces the spin to align along the $z$-direction. Together with the symmetry constraints imposed by $M_x$ and $M_y$, the Bloch eigenstates of this material exhibit vanishing spin expectation values at equilibrium. To induce spin dynamics, these symmetries must be broken by external electric fields. However, unlike the case of monolayer Bi$_2$Se$_3$, an in-plane field alone is insufficient due to the presence of the glide symmetry $\hat G$ or $\hat M_z$. It is necessary to break one of the mirror symmetries ($M_x$ or $M_y$) simultaneously with the glide symmetry. To this end, we apply an out-of-plane electric field $E_z$ together with an in-plane bias along the $y$-direction, as depicted in Fig \ref{Fig2}(b, d). Under the effect of these symmetry-lowering fields, a transverse spin accumulation is induced along the $x$-direction, as presented in Fig. \ref{Fig2}(f).
A key lesson common to the two real-material examples, presented in Fig. \ref{Fig2}, is that spin dynamics, absent in the canonical SOC Hamiltonian ($\hat H _{\text{SOC}}^{\text{CN}}$), sharply emerge when the gauge-field contribution to SOC is included. As already illustrated in Fig. \ref{Fig1}(c) for the Rashba two-band model, the central issue is not the strength of the external field ($\vec E _{\text{ext}}(t)$), but the symmetry-breaking introduced by the SOC gauge-field term. Even a weak field can produce a non-negligible effect as it accumulates over the real-time evolution. Note that the electric field accumulates into the gauge field--the vector potential--over time,
\begin{equation}
    {{\mathbf{A}}_{\text{ext}}}(t) =  - c\int_0^t {{{\mathbf{E}}_{\text{ext}}}(t')} dt'
    \label{Eq10}
\end{equation}
Note that the mechanical momentum entering $\hat H_{{\text{SOC}}}^{{\text{CN + GF}}}$ depends on the gauge field through ${\hat{\vec \pi }} = {\hat{\vec p}} + \frac{e}{c}{{\vec{A}}_{\text{ext}}}(t)$, which implies that even the effect of a weak field can be accumulated, leading to the substantial contribution of the gauge-field term of the SOC. In short, whenever the gauge-field term lowers the symmetry of the canonical SOC, it must be included irrespective of the field strength. The role of a weak field is not negligible--it merely manifests as a delayed but eventually significant response.

\begin{figure}[h]
    \centering
    \includegraphics{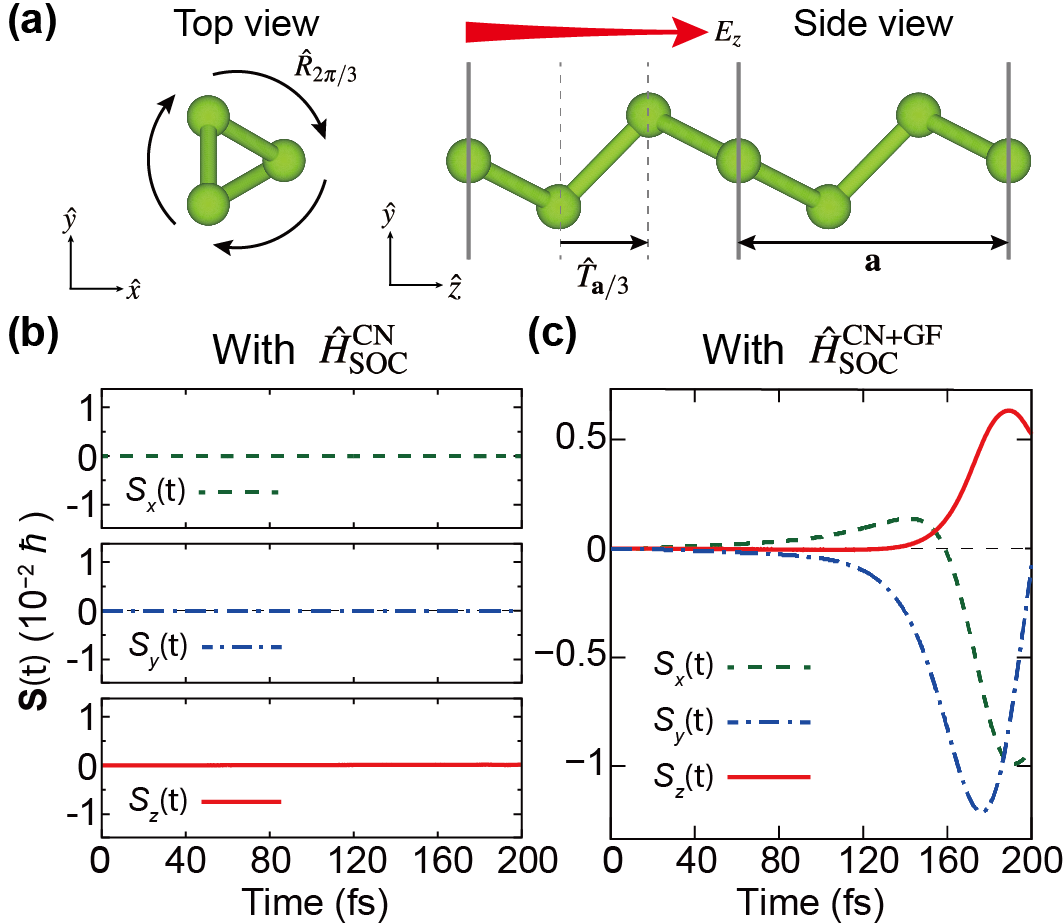}
    \caption{Screw rotational symmetry of a trigonal chiral wire and the symmetry-lowering external axial electric field($E_z \hat z$). (b) Real-time profile of spins obtained with $\hat{H}^{\text{CN}}_{\text{SOC}}$, (c) The same real-time profiles computed with $\hat{H}^{\text{CN+GF}}_{\text{SOC}}$.}
    \label{Fig3}
\end{figure}
In the preceding discussion, we focused on mirror-reflection symmetry. We showed that, even when an external field breaks mirror symmetry, the canonical SOC Hamiltonian $\hat{H}^{\text{CN}}_{\text{SOC}}$ remains intact and continues to constrain the spin dynamics. We now turn to a distinct unitary symmetry, namely the screw-rotational symmetry of a one-dimensional chiral wire. Figure \ref{Fig3}(a) shows the top and side views of a trigonal chiral Se chain\cite{AN2007357, Gates2002, RN8}. This structure possesses a screw-rotational symmetry defined as a combination of partial translation and rotation:
\begin{equation}
    \begin{gathered}
    \hat Q = {{\hat T}_{a/3}}{{\hat R}_{2\pi /3}} = {{\hat R}_{2\pi /3}}{{\hat T}_{a/3}} \\ 
    [\hat H,\hat Q] = 0, \\ 
    \left\langle {{\psi _{n,k}}|{{\hat Q}^{-1} }{\mathbf{\hat S}}\hat Q|{\psi _{n,k}}} \right\rangle  = \left\langle {{\psi _{n,k}}|\hat R_{2\pi /3}^{-1} {\mathbf{\hat S}}{{\hat R}_{2\pi /3}}|{\psi _{n,k}}} \right\rangle  \\ 
    \end{gathered} 
    \label{Eq11}
\end{equation}
In the third line of Eq. (\ref{Eq11}), we used the fact that the spin operator is unaffected by the translational component of $\hat Q$. Consequently, the symmetry enforces that the spin expectation value must be equal to that of a spin rotated by $\hat R_{2 \pi /3}$. This immediately implies that the in-plane spin($S_x$, $S_y$) components vanish. Although $S_z$ is not constrained by the screw-rotational symmetry, the sum over all occupied bands must vanish due to Kramer's pairing enforced by time-reversal symmetry:
\begin{equation}
\begin{gathered}
  \left\langle {{\psi _{n,k}}|{{\hat S}_x}|{\psi _{n,k}}} \right\rangle  = 0 = \left\langle {{\psi _{n,k}}|{{\hat S}_y}|{\psi _{n,k}}} \right\rangle  \\ 
  \left\langle {{\psi _{n,k}}|{{\hat S}_z}|{\psi _{n,k}}} \right\rangle  \ne 0 \\ 
  \int {dk} \left\langle {{\psi _{n,k}}|{{\hat S}_z}|{\psi _{n,k}}} \right\rangle  = 0 \\ 
\end{gathered}
\label{Eq12}
\end{equation}
By applying an external electric field along the screw axis ($z$-direction), we can intentionally break the screw-rotational symmetry as depicted in Fig. \ref{Fig3}(b). Here, we consider the release of the screw-symmetry constraint by an axial electric field ($E_z \hat z$). Consistent with our central argument, when the time evolution is governed solely by $\hat H_{{\text{SOC}}}^{{\text{CN}}}$, the whole spins are well frozen under the symmetry constraints, as shown in Fig. \ref{Fig3}(b). In contrast, inclusion of the gauge-field contribution, $\hat H_{{\text{SOC}}}^{{\text{CN+GF}}}$, leads to clearly observable spin dynamics, as presented in Fig. \ref{Fig3}(c).

\begin{figure}[h]
    \centering
    \includegraphics{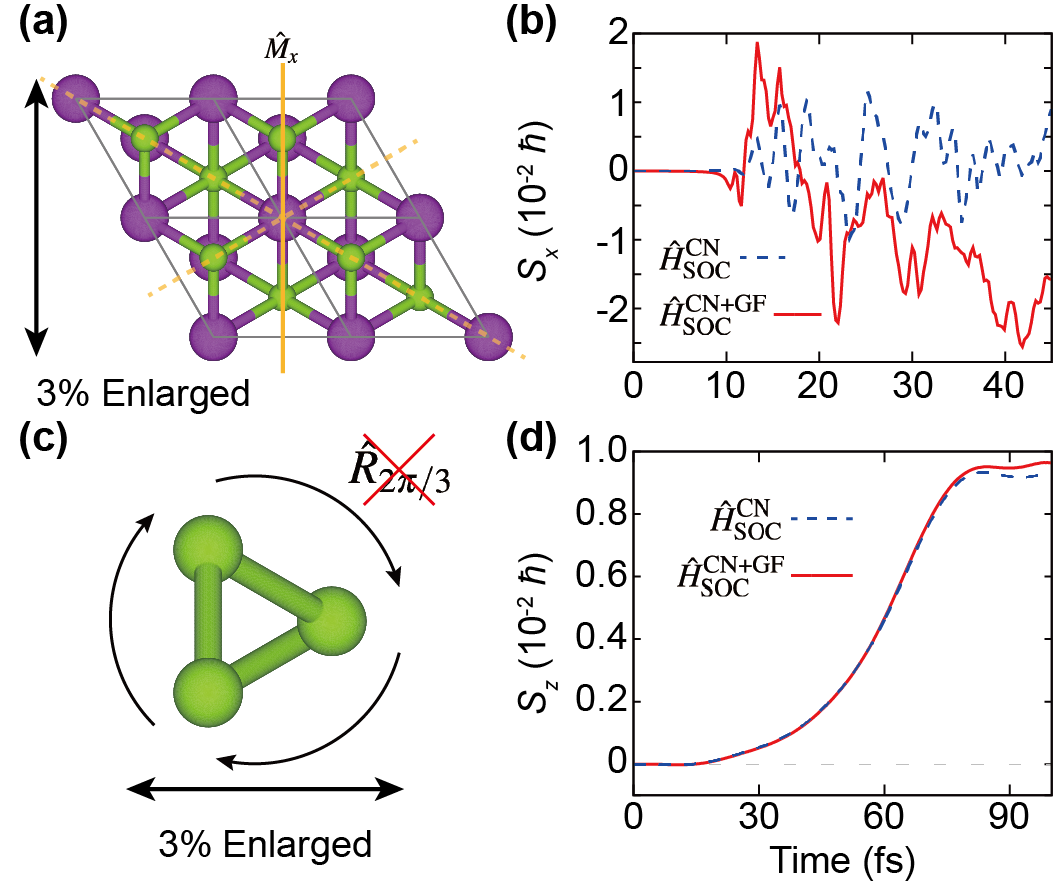}
    \caption{Influence of $\hat{H}^{\text{GF}}_{\text{SOC}}$ on a low-symmetry system for which the gauge-field term does not further break the symmetry. (a) Schematic illustration of Bi$_2$Se$_3$ subjected to a 3\% strain along the $y$ direction, which breaks two mirror symmetries while preserving only ${\hat{M}}_y$. (b) Calculated $x$-direction spin ($S_x$) under the effect of electric bias applied along the $y$ direction. (c) Se trigonal chain with 3\% strain along the $x$-direction, removing the screw-rotation symmetry. (d) Calculated $z$-direction spin ($S_z$) of the strained Se chain in (c) under an electric bias applied along the $z$ direction. In (b) and (d), the red solid lines correspond to the results obtained with the SOC term including the gauge-field term, whereas the blue dashed lines denote results obtained only with the canonical SOC.}
    \label{Fig4}
\end{figure}
In the preceding paragraphs, we emphasized that when an external field lowers the symmetry--such that $\hat{H}^{\text{CN+GF}}_{\text{SOC}}$ possesses a symmetry distinct from $\hat{H}^{\text{CN}}_{\text{SOC}}$, the gauge-field contribution becomes very essential, regardless of whether the external field is strong or weak. We now proceed to quantitatively compare the two formulations of SOC in a situation where the gauge-field term does not further reduce the symmetry beyond that already imposed by $\hat{H}^{\text{CN}}_{\text{SOC}}$, i.e., when the canonical part itself already has reduced symmetry. As a concrete example, as depicted in Fig. \ref{Fig4}(a), we introduce a uniaxial strain along the $y$-direction in otherwise $C_{3v}$-symmetric Bi$_2$Se$_3$, thereby removing the two mirrors, preserving the symmetry to a single mirror plane $\hat{M}_x$. An electric field is then applied along the $y$-direction: $\vec E_{\text{ext}}(t)=E_y(t)\hat y$. We perform rt-TDDFT calculations using both $\hat{H}^{\text{CN}}_{\text{SOC}}$ and $\hat{H}^{\text{CN+GF}}_{\text{SOC}}$. Since both Hamiltonians preserve the single mirror plane($\hat M_x$), the spin expectation value is constrained to develop only along the $x$-direction. The resulting dynamics from the two formulations do not differ significantly. Figure \ref{Fig4}(b) shows the time evolution of the $x$-component of the spin. In addition, we consider a screw-rotationally symmetric trigonal Se chain. By applying a uniaxial strain, as illustrated in Fig. \ref{Fig4}(c), we explicitly break the screw-rotational symmetry. As in the previous case, an electric field is applied along the screw axis. As shown in Fig. \ref{Fig4}(b, d), once the geometrical symmetry is already reduced, the contribution from the gauge-field term in $\hat{H}^{\text{CN+GF}}_{\text{SOC}}$ becomes not that drastic, even negligible.

\section{Methods}\label{sec3}
For the rt-TDDFT calculations in the main text, we first need to obtain the ground state charge density and wavefunctions. These calculations are performed using the Quantum ESPRESSO package\cite{QE1, QE2}. The exchange-correlation energy was treated within the Perdew-Burke-Ernzerhof (PBE) form of the generalized gradient approximation (GGA)\cite{RN5}. Fully relativistic, norm-conserving pseudopotentials were used to account for atomic potentials and spin-orbit coupling. The energy cutoff for the plane-wave basis set is set to 60 Ry. For the k-point sampling to achieve an appropriate electronic structure, we selected 21$\times$21$\times$1 for Bi$_2$Se$_3$, 9$\times$9$\times$5 for ZrSiS, and 1$\times$1$\times$31 for the 1D trigonal Se chain. Using the ground-state eigenvectors obtained, we performed the rt-TDDFT calculation as follows.
\begin{equation}
\begin{gathered}
  i\hbar \frac{\partial }{{\partial t}}{\psi _{n,{\vec{k}}}}({\vec{r}},t) = \hat H({\vec{r}},t){\psi _{n,{\vec{k}}}}({\vec{r}},t) \\ 
  \hat H({\vec{r}},t) = \frac{1}{{2m}}{{{\vec{\hat \pi }}}^2} + \sum\limits_\lambda ^{} {{V_{atom}}} ({\vec{r}} - {{\vec{R}}_\lambda }) + {V_{DFT}}\left[ {\rho ({\vec{r}},t)} \right] + \hat H_{\text{SOC}} \\ 
  {\vec{\hat \pi }} =  - i\hbar \nabla  + \frac{e}{c}{{\vec{A}}_{\text{ext}}}(t) + \frac{im}{\hbar}[\hat V _{NL},\hat {\vec r}]\\ 
  \hat H_{\text{SOC}}^{\text{CN}} = \frac{\hbar}{4m^2c^2} \vec \sigma \cdot \nabla U(\vec r) \times -i \hbar \nabla, \quad \hat H_{\text{SOC}}^{\text{CN+GF}} = \frac{e\hbar}{4m^2c^3} \vec \sigma \cdot \nabla U(\vec r) \times ( -i \hbar \nabla+\frac{e}{c}\vec A_{\text{ext}} (t)) \\
\end{gathered}
\label{Eq13}
\end{equation}
In this equation, ($n, \vec{k}$) denotes the band index and the Bloch wave vector, while $V_{atom}(\vec{r-R}_{\lambda})$ represents the pseudopotential of the $\lambda$-th atom. In the second line in Eq. (\ref{Eq13}), $V_{DFT}\left[ \rho ({\vec{r}},t) \right]$ collectively includes the density-dependent potentials, such as the Hartree and exchange-correlation contributions. In the 4th line of Eq. (\ref{Eq13}), $U(\vec r)$ denotes the total local potential. We introduced the time-dependent vector potential to simulate a spatially uniform electric bias under the velocity gauge condition\cite{Shin2019pnas, yabana2006real, PhysRevB.62.7998, PhysRevB.52.R2225, shin2016real}. For time integration, we employed the Crank–Nicolson method for its numerical stability and computational efficiency. In these simulations, the discretized time step ($\Delta t$) was set to 2.414 attoseconds. 

\section{Discussion}\label{sec4}
We investigated spin dynamics in spin–orbit–coupled systems under external electromagnetic fields and showed that the gauge-field contribution to SOC is essentially required and cannot be ignored, particularly for nonequilibrium dynamical states. Using both a two-band Rashba model and first-principles calculations of real materials with mirror, glide, or screw symmetries, we found that once an external field breaks the underlying symmetry, the gauge-field term becomes quantitatively significant even for a weak external field. Our results demonstrate that a fully gauge-invariant formulation of SOC is not limited to a formal requirement, but is necessary for accurate predictions of any nonequilibrium spin dynamics. It should be noted that, in the velocity gauge, a DC bias drives the vector potential to grow monotonically in time, making the gauge-field contribution to the Hamiltonian increasingly pronounced--unlike the AC case, where the oscillatory nature of the vector potential renders this effect comparatively negligible.

\bmhead{Supplementary information}
Supplementary Information accompanies this paper. Supplementary Note 1 provides a detailed derivation of the gauge-field-modified Kleinman–Bylander projectors within the velocity-gauge pseudopotential formalism. Supplementary Figs. 1-5 show the spin-resolved band structure of the two-dimensional Rashba square-lattice model(Fig. 1), the crystal structure of ZrSiS illustrating the nonsymmorphic glide symmetry(Fig. 2), the electronic band structures of Bi$_2$Se$_3$ and ZrSiS used as initial wavefunctions(Fig. 3), and the full three-component real-time spin dynamics of Bi$_2$Se$_3$(Fig. 4) and ZrSiS(Fig. 5), respectively.

\bmhead{Acknowledgements}
This work was supported by the National Research Foundation of Korea(NRF) grant funded by the Korea government(MSIT) (No. RS-2023-00257666, No.RS-2023-00208825) and Korea Institute for Advancement of Technology(KIAT) grant funded by the Korea Government(MOTIE)(P0023703, HRD Program for Industrial Innovation). 

\begin{itemize}
\item Conflict of interest/Competing interests 

The authors declare no competing interests.
\item Data availability 

The data supporting the findings of this article are not publicly available upon publication because it is not technically feasible, or the cost of preparing, depositing, and hosting the data would be prohibitive within the terms of this research project. The data are available from the authors upon reasonable request.
\item Author contribution

U.J. conceived the theoretical framework, performed all rt-TDDFT calculations, and drafted the manuscript. E.T.S., A.R., and C.A.U. provided theoretical guidance and critically revised the manuscript. K.-W.K. and N.P. supervised the project, directed the scientific interpretation, and revised the manuscript. All authors discussed the results and approved the final version.
\end{itemize}

\bibliography{sn-bibliography}% common bib file

\end{document}

\title[Article Title]{Supplementary Information of Gauge-Field-Mediated Symmetry Breaking of Matters Under Electromagnetic Fields and Its Impact on Spin Dynamics}


\author[1]{\fnm{Uiseok} \sur{Jeong}}\email{uiseok95@unist.ac.kr}
\author[1]{\fnm{Esmaeil} \sur{Taghizadeh Sisakht}} \email{sisakht@unist.ac.kr}
\author[2]{\fnm{Angel} \sur{Rubio}} \email{angel.rubio@mpsd.mpg.de}
\author[3]{\fnm{Carsten} \sur{A. Ullrich}} \email{ullrichc@missouri.edu}
\author*[4]{\fnm{Kyoung-Whan} \sur{Kim}}\email{kwkim@yonsei.ac.kr}
\author*[1,2]{\fnm{Noejung} \sur{Park}}\email{noejung@unist.ac.kr}

\affil[1]{\orgdiv{Department of Physics}, \orgname{Ulsan National Institute of Science and Technology}, \orgaddress{\street{50, UNIST-gil}, \city{Ulju-gun}, \postcode{44919}, \state{Ulsan}, \country{Republic of Korea}}}

\affil[2]{\orgdiv{Department of Physics}, \orgname{Max Planck Institute for the Structure and Dynamics of Matter}, \orgaddress{\street{Luruper Ch 149}, \city{Hamburg}, \postcode{22607}, \country{Germany}}}

\affil[3]{\orgdiv{Department of Physics and Astronomy}, \orgname{University of Missouri}, \orgaddress{\city{Columbia}, \postcode{65211}, \state{Missouri}, \country{USA}}}

\affil[4]{\orgdiv{Department of Physics}, \orgname{Yeonsei University}, \orgaddress{\street{50, Yonsei-ro}, \city{Seodaemun-gu}, \postcode{03722}, \state{Seoul}, \country{Republic of Korea}}}


\abstract{When a condensed-matter system is subjected to external electromagnetic fields, the gauge-invariant formulation of physical operators must explicitly incorporate the gauge-field contribution. However, in the context of spin–orbit coupling (SOC), this gauge-field term is often regarded as negligible or merely additive compared to the canonical SOC, which is typically localized near atomic cores. Here, we demonstrate that the symmetry breaking and consequent spin dynamics are governed by the gauge-field term, without which the spins remain symmetry-constrained. We perform real-time time-dependent density functional theory calculations to investigate spin–orbit dynamics, focusing on representative cases with mirror, glide, and screw-rotational symmetry. We demonstrate that when the gauge-field term in the time-dependent Hamiltonian perturbs the symmetry of the canonical term, a dynamical spin state gradually develops during the time evolution, beyond the symmetry-frozen states. We suggest that, for nonequilibrium spin–orbit dynamics, the gauge-invariant formulation of SOC is not only formally required but also quantitatively essential, even for a weak external field.}

\keywords{Spin-Orbit Coupling, Spin dynamics, Spintronics, rt-TDDFT}

\maketitle

Supplementary Note 1---Here, we describe how the gauge-field term of SOC (arising from the external vector potential) is incorporated within the Kleinman–Bylander nonlocal projector scheme. Let us start with the time-dependent equation in the velocity gauge with the external vector potential.
\begin{equation}
    i\hbar \frac{\partial }{{\partial t}}\Psi  = \left[ {\frac{1}{{2m}}{{\left( { - i\hbar \vec \nabla  + \frac{e}{c}{{\mathbf{A}}_{\text{ext}}}} \right)}^2} + U({\mathbf{r}}) + \frac{\hbar }{{4{m^2}{c^2}}}{\mathbf{\sigma }} \cdot \vec \nabla U \times \left( { - i\hbar \vec \nabla  + \frac{e}{c}{{\mathbf{A}}_{\text{ext}}}} \right)} \right]\Psi
    \label{EqS1}
\end{equation}
In what follows, we specify the case of a uniform electric field, but similar formalism can be naturally extended to non-uniform cases. To be fit into the standard pseudopotential technique, we need to absorb the vector potential onto the $U(1)$ phase, and the right-hand side of Eq. (\ref{EqS1}) can be rephrased as 
\begin{equation}
\left[ {\frac{1}{{2m}}{{\left( { - i\hbar \vec \nabla  + \frac{e}{c}{{\mathbf{A}}_{\text{ext}}}} \right)}^2} + U({\mathbf{r}})} \right]\Psi  + {e^{ - i{{\mathbf{A}}_{\text{ext}}} \cdot {\mathbf{r}}/\hbar c}}\left[ {\frac{\hbar }{{4{m^2}{c^2}}}{\mathbf{\sigma }} \cdot \vec \nabla U \times ( - i\hbar \vec \nabla )} \right]{e^{i{{\mathbf{A}}_{\text{ext}}} \cdot {\mathbf{r}}/\hbar c}}\Psi.
\label{EqS2}
\end{equation}
The same treatment can be given to the Bloch wave-vector part, and Eq. (\ref{EqS1}) can be rewritten in terms of the cell periodic part of the Bloch wave.
\begin{equation}
\begin{gathered}
  \Psi ({\mathbf{r}},t) = {e^{i{\mathbf{k}} \cdot {\mathbf{r}}}}\chi ({\mathbf{r}},t) \\ 
  i\hbar \frac{\partial }{{\partial t}}\chi ({\mathbf{r}},t) = \left[ \begin{gathered}
  \frac{1}{{2m}}{\left( { - i\hbar \nabla  + \hbar {\mathbf{k}} + \frac{e}{c}{{\mathbf{A}}_{ext}}} \right)^2} + U({\mathbf{r}}) \hfill \\
   + \exp \left( { - i(\frac{{{{\mathbf{A}}_{ext}} \cdot {\mathbf{r}}}}{{\hbar c}} + {\mathbf{k}})} \right)\frac{\hbar }{{4{m^2}{c^2}}}{\mathbf{\sigma }} \cdot \nabla U \times ( - i\hbar \nabla )\exp \left( {i(\frac{{{{\mathbf{A}}_{ext}} \cdot {\mathbf{r}}}}{{\hbar c}} + {\mathbf{k}})} \right) \hfill \\ 
\end{gathered}  \right]\chi ({\mathbf{r}},t) \\ 
\end{gathered}
\label{EqS3}
\end{equation}
In the standard pseudopotential method, the spin-orbit coupling terms, together with the non-local part of the pseudopotential, are counted by angular momentum partial waves at each atomic core, for which the Kleinmann-Bylander type fully separable form is constructed. For instance, 
\begin{eqnarray}
\frac{\hbar }{{4{m^2}{c^2}}}{\mathbf{\sigma }} \cdot \vec \nabla U \times ( - i\hbar \vec \nabla ) \simeq&& \nonumber\\
\sum\limits_\tau ^{Natom}&& {\sum\limits_{j,m} {\left| {{\beta _{\tau ,j,m}}} \right\rangle } } \left\langle {{\beta _{\tau ,j,m}}} \right|.
    \label{EqS4}
\end{eqnarray}
The phase factor given in Eq. (\ref{EqS3}) will be counted by redefinition of the projectors, as follows:
\begin{equation}
    \begin{gathered}
    \exp \left( { - i(\frac{{{{\mathbf{A}}_{\text{ext}}} \cdot {\mathbf{r}}}}{{\hbar c}} + {\mathbf{k}})} \right){\text{ }}\left[ {\frac{\hbar }{{4{m^2}{c^2}}}{\mathbf{\sigma }} \cdot \vec \nabla U \times ( - i\hbar \vec \nabla )} \right]\exp \left( {i(\frac{{{{\mathbf{A}}_{\text{ext}}} \cdot {\mathbf{r}}}}{{\hbar c}} + {\mathbf{k}})} \right) \\
    \simeq \sum\limits_\tau ^{Natom} {\sum\limits_{j,m} {\left| {{{\tilde \beta }_{\tau ,j,m}}} \right\rangle } } \left\langle {{{\tilde \beta }_{\tau ,j,m}}} \right|.    
    \end{gathered}
    \label{EqS5}
\end{equation}
The position-space representation of the redefined projector can now be written as
\begin{eqnarray}
\left\langle {{\mathbf{r}}\left| {{{\tilde \beta }_{\tau ,j,m}}} \right.} \right\rangle  = {e^{-i{{\mathbf{A}}_{\text{ext}}} \cdot {\mathbf{r}}/\hbar c}}{e^{-i{\mathbf{k}} \cdot {\mathbf{r}}}}\left\langle {{\mathbf{r}}\left| {{\beta _{\tau ,j,m}}} \right.} \right\rangle \nonumber \\ = {e^{-i{{\mathbf{A}}_{\text{ext}}} \cdot {\mathbf{r}}/\hbar c}}{e^{-i{\mathbf{k}} \cdot {\mathbf{r}}}}{\beta _{\tau ,j,m}}({\mathbf{r}}).
    \label{EqS6}
\end{eqnarray}
The velocity-gauge wave function, given in Eq. (\ref{EqS1}) will be expanded in terms of plane waves,
\begin{equation}
\begin{gathered}
    \left| {\chi_{n,\mathbf{k}} (t)} \right\rangle  = \sum\limits_{\mathbf{G}} {A_{n,\mathbf{k}}({\mathbf{G}},t} )\left| {\mathbf{G}} \right\rangle {\text{,}} \\ 
\chi_{n,\mathbf{k}} ({\mathbf{r}},t) = \sum\limits_{\mathbf{G}} {A_{n,\mathbf{k}}({\mathbf{G}},t} )\left\langle {\mathbf{r}} \right.\left| {\mathbf{G}} \right\rangle = \sum\limits_{\mathbf{G}} {A_{n,\mathbf{k}}({\mathbf{G}}} ,t)\frac{1}{{\sqrt \Omega  }}{e^{i{\mathbf{G}} \cdot {\mathbf{r}}}}
\end{gathered}
\label{EqS7}
\end{equation}
 
Then, the time-dependent equation Eq. (\ref{EqS1}) for the Bloch wave is now given by 
\begin{equation}
\begin{gathered}
    i\hbar \frac{\partial }{{\partial t}}A_{n,\mathbf{k}}({\mathbf{G}},t) = \sum\limits_{{\mathbf{G'}}} {\left[ {\frac{1}{{2m}}{{(\hbar {\mathbf{G'}} + \hbar {\mathbf{k}+\frac{e}{c}\mathbf{A}_{\text{ext}}(t)})}^2}{\delta _{{\mathbf{G}},{\mathbf{G'}}}} + {U_{local}}({\mathbf{G}} - {\mathbf{G'}})} \right]} A_{n,\mathbf{k}}({\mathbf{G'}},t) + \\
    \sum\limits_{\tau ,l,m} {\left\langle {{\mathbf{G}}\left| {{{\tilde \beta }_{\tau ,l,m}}} \right.} \right\rangle \left\langle {{{\tilde \beta }_{\tau ,l,m}}\left| {{\chi_{n,{\mathbf{k}}}}} \right.} \right\rangle }    
\end{gathered}
    \label{EqS8}
\end{equation}
The projector's action on the wavefunction can be calculated in either position or momentum space. 
\begin{equation}
\begin{gathered}
  \left\langle {{{\tilde \beta }_{\tau ,l,m}}\left| {{\chi_{n,{\mathbf{k}}}}} \right.} \right\rangle  = \sum\limits_{\mathbf{G}} {\left\langle {{{\tilde \beta }_{\tau ,l,m}}\left| {\mathbf{G}} \right.} \right\rangle } A_{n,\mathbf{k}}({\mathbf{G}}) = \sum\limits_{\mathbf{G}} {\tilde \beta _{\tau ,l,m}^*({\mathbf{G}})} A_{n,\mathbf{k}}({\mathbf{G}}) \\ 
   = \int {{d^3}} {\mathbf{r}}\tilde \beta _{\tau ,l,m}^*({\mathbf{r}}){\chi_{n,{\mathbf{k}}}}({\mathbf{r}}) = \int {{d^3}} {\mathbf{r}}{e^{i{{\mathbf{A}}_{\text{ext}}} \cdot {\mathbf{r}}/\hbar c}}{e^{i{\mathbf{k}} \cdot {\mathbf{r}}}}\beta _{\tau ,l,m}^*({\mathbf{r}}){\chi_{n,{\mathbf{k}}}}({\mathbf{r}}) \\ 
\end{gathered}
\label{EqS9}
\end{equation}
G-component of the projectors are defined as follows. 
\begin{equation}
\begin{gathered}
  {\beta _{\tau ,j,m}}({\mathbf{r}} - \vec \tau ) = \sum\limits_{\mathbf{G}} {{\beta _{\tau ,j,m}}({\mathbf{G}}){e^{i{\mathbf{G}} \cdot {\mathbf{r}}}}}  \\ 
  \left\langle {{\mathbf{G}}\left| {{\beta _{\tau ,j,m}}} \right.} \right\rangle  = {\beta _{\tau ,j,m}}({\mathbf{G}}) = \frac{1}{{{V_{cell}}}}\int {{\beta _{\tau ,j,m}}({\mathbf{r}} - \vec \tau )} {e^{ - i{\mathbf{G}} \cdot {\mathbf{r}}}} \\
  = {e^{ - i{\mathbf{G}} \cdot \vec \tau }}\frac{1}{{{V_{cell}}}}\int {{\beta _{\tau ,j,m}}(r)} {e^{ - i{\mathbf{G}} \cdot {\mathbf{r}}}}{d^3}{\mathbf{r}} \\ 
  \left\langle {{\mathbf{G}}\left| {{{\tilde \beta }_{\tau ,j,m}}} \right.} \right\rangle  = \frac{1}{{{V_{cell}}}}\int {{\beta _{\tau ,j,m}}({\mathbf{r}} - \vec \tau )} \exp \left( { - i(\frac{{{{\mathbf{A}}_{ext}}}}{{\hbar c}} + {\mathbf{k}} + {\mathbf{G}}) \cdot {\mathbf{r}}} \right){d^3}{\mathbf{r}} \\
  = {e^{ - i{\mathbf{G}} \cdot \vec \tau }}\frac{1}{{{V_{cell}}}}\int {{\beta _{\tau ,j,m}}(r)} {e^{ - i{\mathbf{q}} \cdot {\mathbf{r}}}}{d^3}{\mathbf{r}} \\ 
\end{gathered}
\label{EqS10}
\end{equation}
Here, $\mathbf{\tau}$ denotes the position of the $\tau$-th atom at which the nonlocal projector is centered, and ${\mathbf{q}} = \frac{{{{\mathbf{A}}_{ext}}}}{{\hbar c}} + {\mathbf{k}} + {\mathbf{G}}$ is the time-dependent effective momentum shift introduced by the gauge field. Thus, we can utilize the expansion of the planes in terms of spherical Bessel functions and spherical Harmonics.
\begin{equation}
\begin{gathered}
  {e^{ - i{\mathbf{q}} \cdot {\mathbf{r}}}} = 4\pi \sum\limits_{l = 0}^\infty  {\sum\limits_{m =  - l}^l {{i^l}} } {j_l}(qr)Y_{lm}^*(\hat q){Y_{lm}}(\hat r){\text{  ,  }}{\mathbf{q}} = \frac{{{{\mathbf{A}}_{ext}}}}{{\hbar c}} + {\mathbf{k}} + {\mathbf{G}} \\ 
  \left\langle {{\mathbf{G}}\left| {{{\tilde \beta }_{\tau ,j,m}}} \right.} \right\rangle  = 4\pi {e^{ - i{\mathbf{G}} \cdot \vec \tau }}Y_{lm}^*(\hat q)\frac{(-i)^l}{{{V_{cell}}}}\int\limits_0^\infty  {{\beta _{\tau ,j,m}}(r){r^2}} {j_l}(qr)dr \\ 
\end{gathered}
\label{EqS11}
\end{equation}

\newpage

\begin{figure}
    \includegraphics[width=1\textwidth]{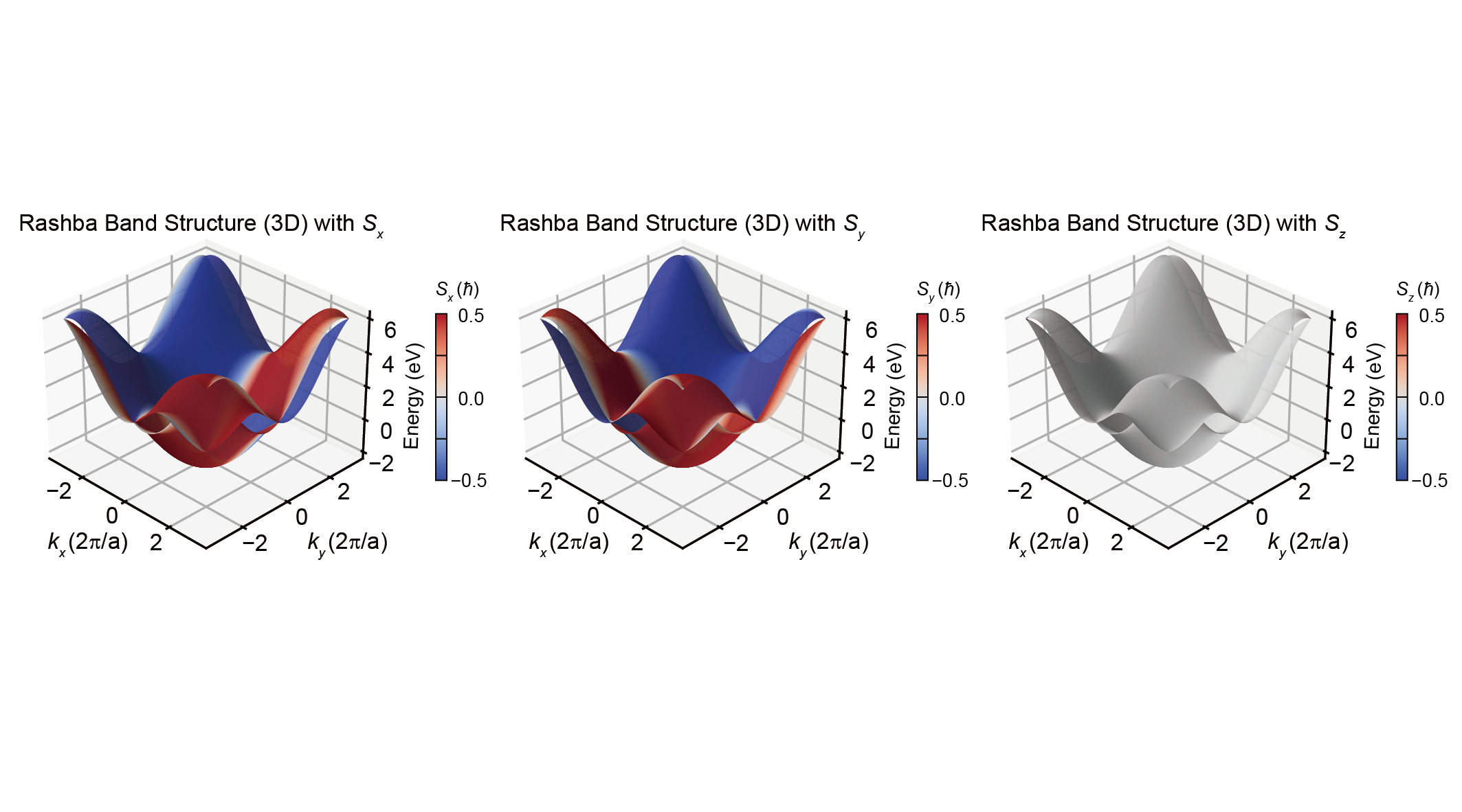}
    \caption{Spin-resolved three-dimensional band structure of the two-dimensional Rashba square-lattice model defined in Eq. (4) of the main text. The energy dispersion $E(\mathbf{k})$ is plotted over the full two-dimensional Brillouin zone, with the color representing the expectation value of each spin component: (left) $S_x$, (center) $S_y$, and (right) $S_z$, in units of $\hbar$.}
    \label{fig:placeholder}
\end{figure}

\begin{figure}
    \includegraphics[width=1\textwidth]{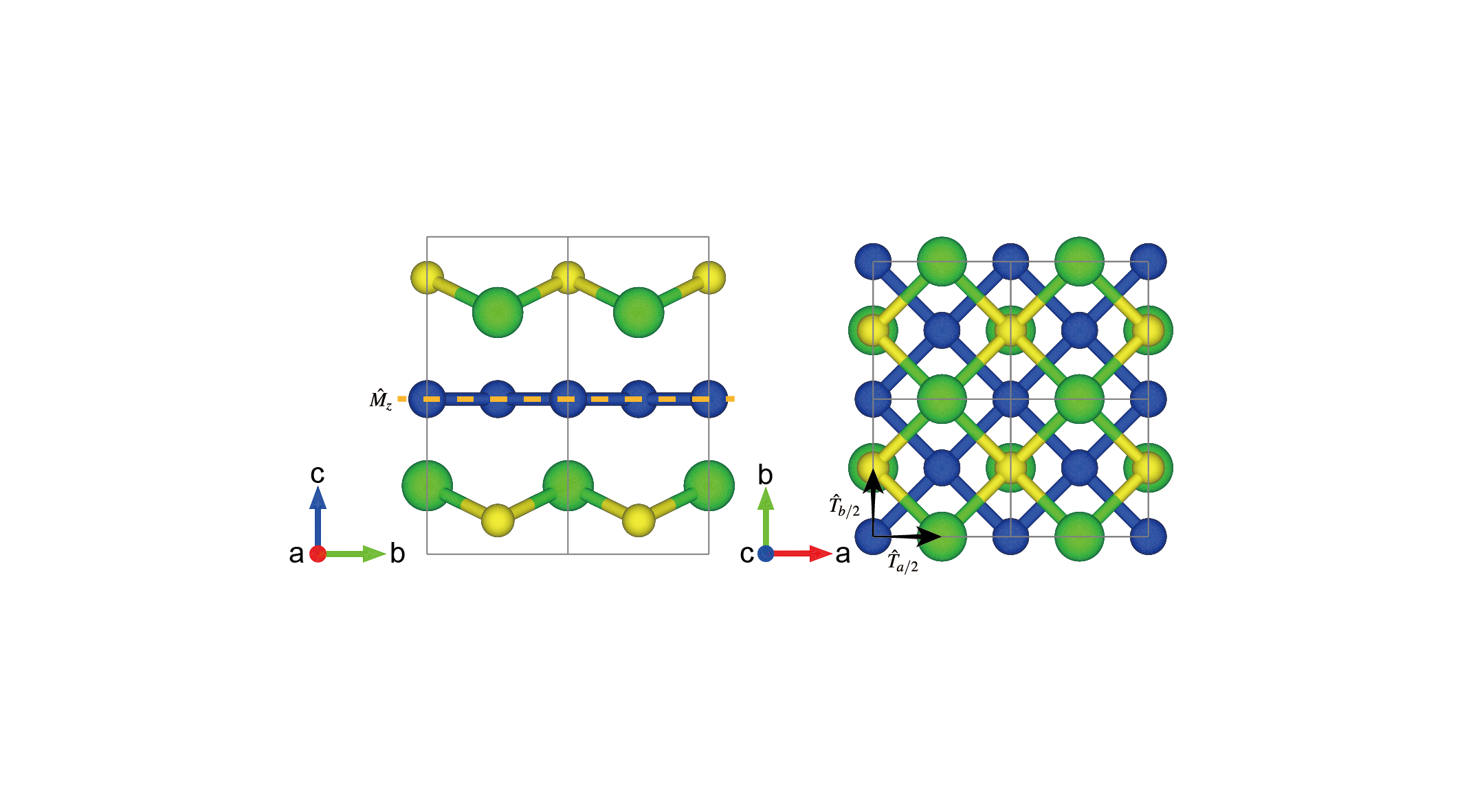}
    \caption{Crystal structure of ZrSiS viewed along the (left) top view and (right) side view, illustrating the nonsymmorphic glide symmetry $\hat{G} = \hat{T}({a}/{4} + {b}/{4})\hat{M}_z $. Blue, green, and yellow spheres represent Zr, Si, and S atoms, respectively. The fractional translation vectors $\hat{t}_{a/2}$ and $\hat{t}_{b/2}$ associated with the glide operation are indicated by arrows. The glide plane, combined with the two mirror symmetries $\hat{M}_x$ and $\hat{M}_y$, enforces all the spins to vanish at equilibrium, as discussed in the main text.}
    \label{fig:placeholder}
\end{figure}

\begin{figure}
    \includegraphics[width=1\textwidth]{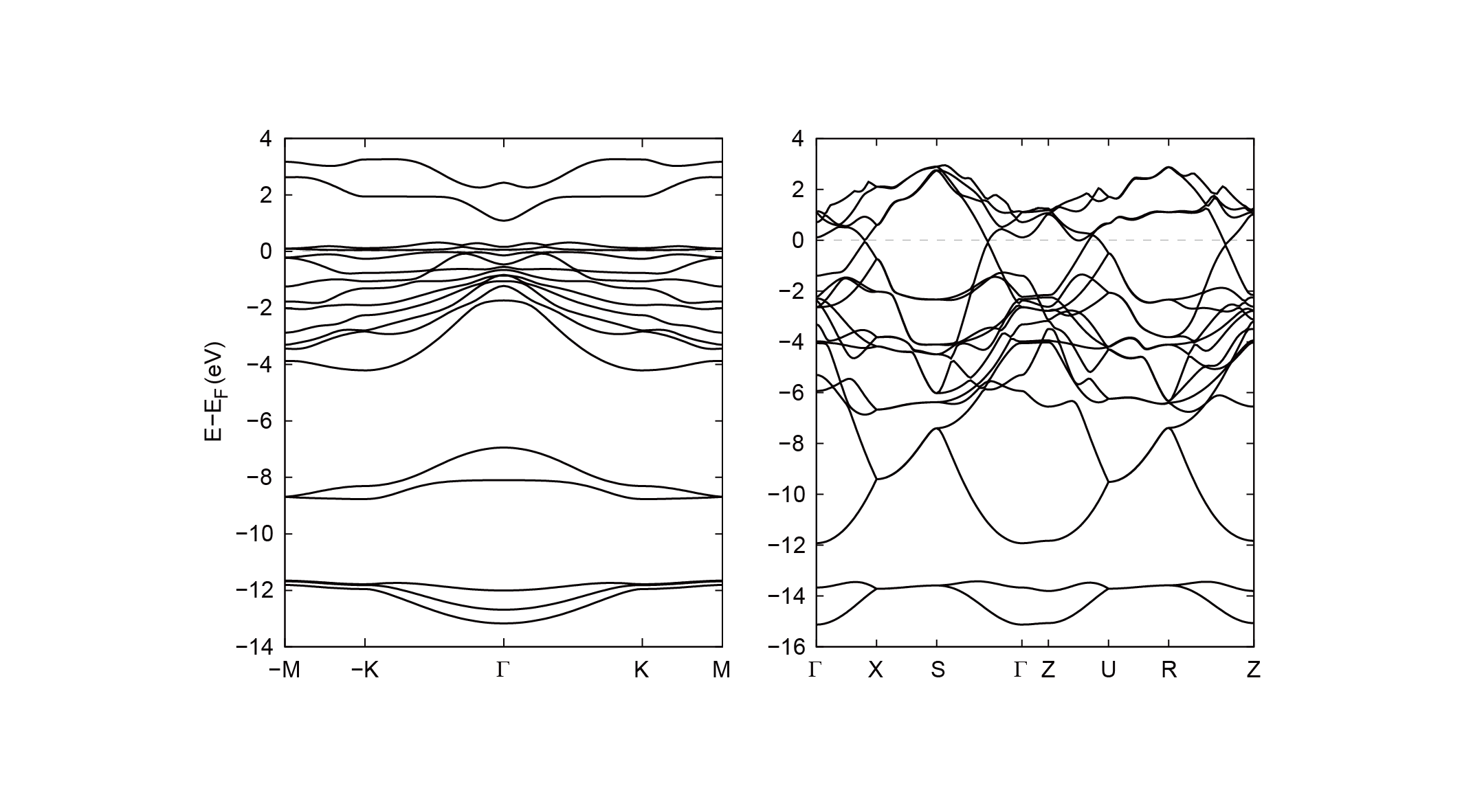}
    \caption{Electronic band structures of Bi$_2$Se$_3$ (left) and ZrSiS (right) used as initial starting wavefunctions for the rt-TDDFT calculations presented in the main text. All energies are referenced to the Fermi level $E_F$.}
    \label{fig:placeholder}
\end{figure}

\begin{figure}
    \includegraphics[width=1\textwidth]{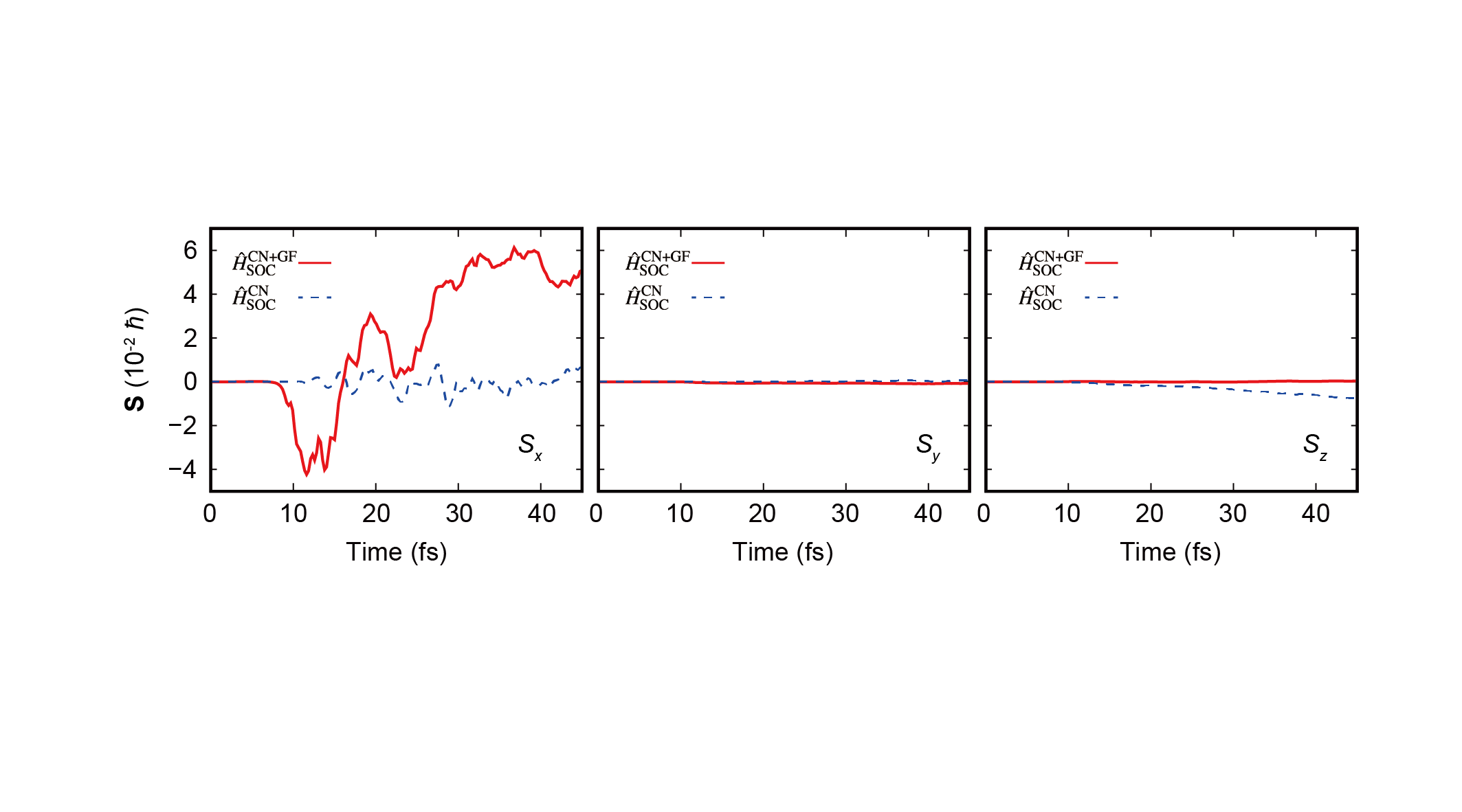}
    \caption{Real-time spin dynamics of Bi$_2$Se$_3$ obtained from rt-TDDFT calculations, showing all three Cartesian components of the total spin expectation value $\mathbf{S}(t)$ (in units of $10^{-2}\hbar$) as a function of time. Results are compared between two formulations of the SOC Hamiltonian: $\hat{H}^\text{CN+GF}_\text{SOC}$ (red solid line), which incorporates both the canonical and gauge-field contributions, and $\hat{H}^\text{CN}_\text{SOC}$ (blue dashed line), which includes only the canonical term. An in-plane electric bias is applied along the y-direction to break the mirror symmetry $\hat{M}_y$ while leaving $\hat{M}_x$ intact, thereby inducing a transverse spin accumulation along the x-direction.}
    \label{fig:placeholder}
\end{figure}

\begin{figure}
    \includegraphics[width=1\textwidth]{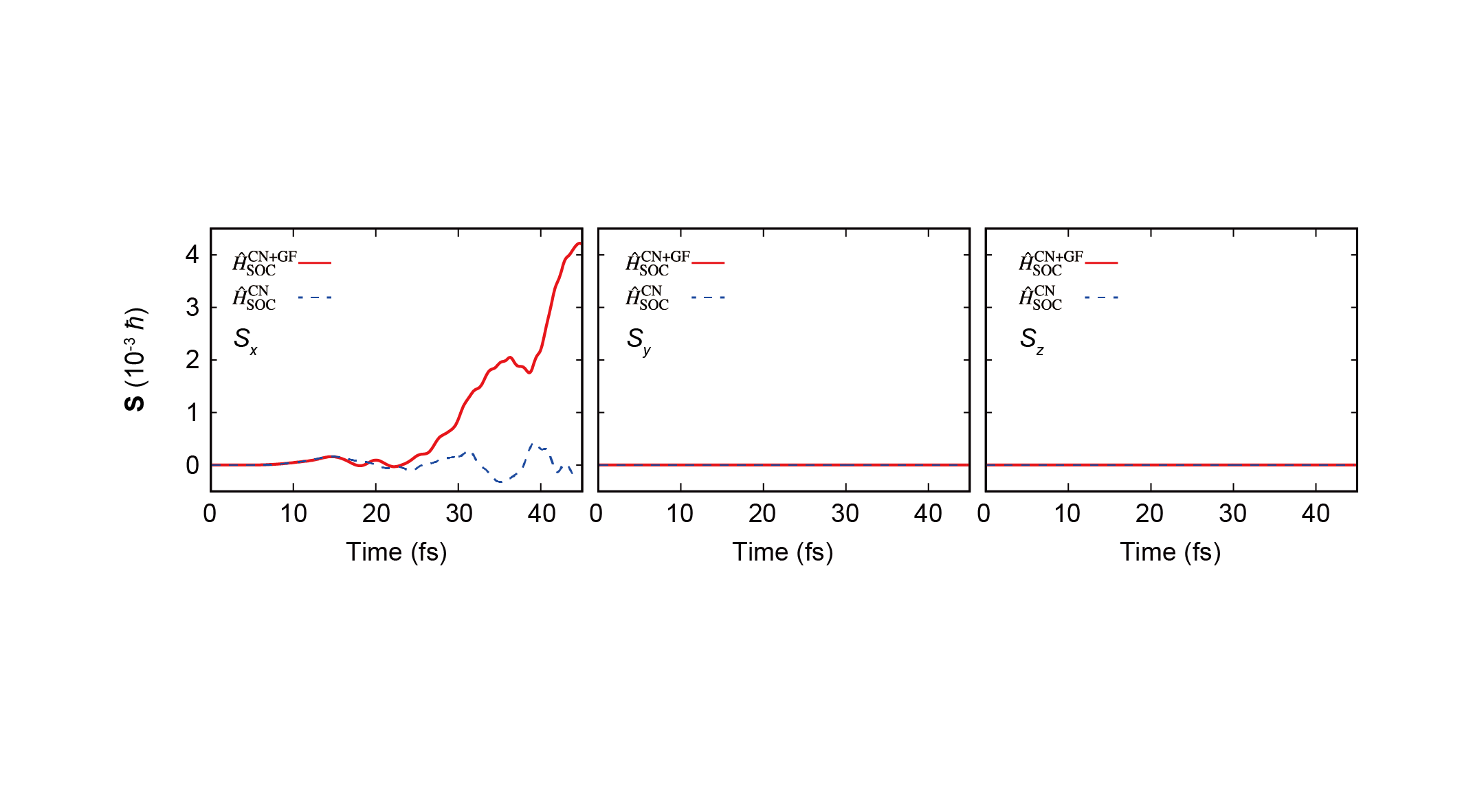}
    \caption{Real-time spin dynamics of ZrSiS obtained from rt-TDDFT calculations, showing all three Cartesian components of the total spin expectation value $\mathbf{S}(t)$ (in units of $10^{-3}\hbar$) as a function of time. Results are compared between two formulations of the SOC Hamiltonian: $\hat{H}^\text{CN+GF}_\text{SOC}$ (red solid line), which incorporates both the canonical and gauge-field contributions, and $\hat{H}^\text{CN}_\text{SOC}$ (blue dashed line), which includes only the canonical term. The simultaneous application of an out-of-plane field $E_z$ and an in-plane bias $E_y$ is required to break both the glide symmetry $\hat{G}$ and the relevant mirror symmetries concurrently.}
    \label{fig:placeholder}
\end{figure}